\documentclass[a4paper, 11pt]{article}
\usepackage{latexsym}
\usepackage{graphicx}
\usepackage{mathptmx}
\usepackage{amsmath}
\usepackage{amsfonts}
\usepackage{amssymb}
\usepackage{amsbsy}
\usepackage{amsthm}%% LaTeX - Article customise

\usepackage{booktabs} % for much better looking tables
\usepackage{array} % for better arrays (eg matrices) in maths
\usepackage{paralist} % very flexible & customisable lists (eg. enumerate/itemize, etc.)
\usepackage{verbatim} % adds environment for commenting out blocks of text & for better verbatim
\usepackage{subfigure} % make it possible to include more than one captioned figure/table in a single float
\usepackage{geometry} % to change the page dimensions
\usepackage{natbib}

\usepackage[pdftex,colorlinks=true,urlcolor=blue,citecolor=black,anchorcolor=black,linkcolor=black]{hyperref}

\begin{document}

%\title{UNIFIED METHOD FOR MARKOV CHAIN TRANSITION MODEL ESTIMATION USING INCOMPLETE SURVEY DATA}
\title{Unified Method for Markov Chain Transition Model Estimation Using Incomplete Survey Data}

% AUTHOR: Enter the authors of the article, see end of the example document for further examples
\author{Duncan Ermini Leaf\\ 
Leonard D. Schaeffer Center for Health Policy and Economics \\
University of Southern California\\
635 Downey Way\\
Los Angeles, CA, 90089-3333, USA}

\maketitle

\section*{ABSTRACT}

The Future Elderly Model and related microsimulations are modeled as Markov chains.
These simulations rely on longitudinal survey data to estimate their transition models.
The use of survey data presents several incomplete data problems, including coarse and irregular spacing of interviews, data collection from subsamples, and structural changes to surveys over time.  
The Expectation-Maximization algorithm is adapted to create a method for maximum likelihood estimation of Markov chain transition models using incomplete data.
The method is demonstrated on a simplified version of the Future Elderly Model.

\section{INTRODUCTION}
\label{sec:intro}

\subsection{Health Policy Simulations}
\label{microsims}
The Future Elderly Model (FEM; \citeauthor*{femtech}, \citeyear{femtech}) is a microsimulation of health-related outcomes for the U.S. population over age 50.  
%[TODO cite some FEM applications, e.g., NAS].
The FEM was recently extended to simulate outcomes for the U.S. population starting at age 25 in the Future Americans Model (FAM; \citeauthor*{famtech}, \citeyear{famtech}).
%[TODO cite some FAM applications, e.g., ABC]
Currently, an international effort is underway to adapt the FEM to populations around the world, including Canada \citep*{compas}, 10 European countries \citep*{eufem}, Japan \citep*{japanfem}, Mexico, Singapore, and South Korea.
Each of these models is simulated as a Markov chain and the transition probabilities are estimated using data from a longitudinal survey in which each individual is typically followed until death.
The FEM is estimated using data from the \citeauthor*{hrs} (\citeyear{hrs}; HRS) core interviews.
Although the specific survey questions differ from country to country, many of these surveys are structured similarly to the HRS and include questions about individuals' health, finances, family structure, utilization of medical services, employment, and participation in government programs \citep*{schs,klosa,mhas,jstar,nphs,share}.

\subsection{Model Specification, Estimation, Simulation, and Analysis}
Policy simulation projects can be divided into three conceptual phases: estimation, simulation, and analysis.  The estimation phase starts with vectors of observed outcomes, $X_t,$ starting at time $t=1$ and ending at $t=T.$  The observations $X_1 \ldots X_T$ are used to estimate the first-order Markov transition probability:
$p(X_t \mid X_{t-1}).$
This is the probability of transitioning into state $X_t$ at time $t$ conditional on being is state $X_{t-1}$ in the previous time step, $t-1$. 
A stationary, first-order Markov chain is assumed here in order to keep notation simple.  Neither of these assumptions are required and it is straightforward to extend the methodology to higher-order or non-stationary chains.
This article focuses on maximum likelihood estimation of a parametric model where the transition probabilities are indexed by a parameter vector $\beta$ so that the transition probability functions have the form $p_t(X_t \mid X_{t-1}; \beta).$  The estimation phase then consists of finding the parameter estimate, $\hat{\beta}$, that maximizes the likelihood of the observed sequence.

Next comes the simulation phase in which Monte Carlo methods are used to simulate $M$ future sequences of outcomes from the estimated transition probability model up to some time $t=T^*$. Formally, for $m = 1, \ldots, M,$
\[
\begin{array}{c}
X_{T+1,m} \sim p(X_{T+1} \mid X_{T,m}; \hat{\beta})\\
X_{T+2,m} \sim p(X_{T+2} \mid X_{T+1, m}; \hat{\beta})\\
\vdots\\
X_{T^*, m} \sim p(X_{T^*} \mid X_{T^*-1, m}; \hat{\beta})
\end{array}
\]

Finally, the simulated outcomes are used as inputs to the analysis phase.  In the health policy setting, simulation output is often used for forecasting, e.g., projecting future medical costs.  The transition model can also be modified to represent a policy intervention and then a second simulation is run to asses the impact of the intervention. For example, a new medical intervention could be represented by reducing the probability of a simulated individual being diagnosed with cancer.  The simulation analyst might look at the average difference in costs between the modified model and the scenario with no intervention to better understand the cost effectiveness of the intervention.

\subsection{Estimation Challenges Stemming from Incomplete Survey Data}
\label{sec:incdata}

The FEM and related simulation projects use observed outcomes, $X_1 \ldots X_T$, from survey data to estimate their transition models.  This reliance on survey data introduces some \emph{incomplete data} problems into the estimation phase.  In the context of this article, \emph{incomplete} means that the survey is missing data on some of the outcomes at some of the time points in the transition model to be estimated. 
Examples include:\\

\noindent \textbf{Coarse spacing of survey waves}: The FEM relies on HRS for transition data.  The FAM uses the \citeauthor*{psid} (\citeyear{psid}; PSID).  In both of these surveys, respondents are interviewed every two years.  For a simulation of one-year time steps, the surveys are missing all outcomes at every other time step.  Only $X_1$, $X_3$, $X_5, \ldots$ are observed\\

\noindent \textbf{Irregular spacing of survey waves}: The Mexico FEM uses \citep*{mhas} for transition data.  Respondents were interviewed in 2001, 2003, 2012, and 2015.  Even if two-year time steps are chosen for simulation, there are missing data for outcomes between 2003 and 2012 and between 2012 and 2015.  Similarly, the FEM for Europe uses the \citep*{share}, which has a two-year wave cycle, but the 2008 survey does not ask the same questions as the other waves and so cannot be used as transition data.\\

\noindent \textbf{Irregular spacing of individual interviews}: There are occasions when a survey cannot reach all of its respondents within a wave cycle due budget or staff limitations.  In these cases the time between a respondent's consecutive interviews will be longer than the wave cycle.  The responses might appear in the survey data as if they were collected within the correct wave, but in fact they should be counted in a later year, which leaves a gap of missing data.\\

\noindent\textbf{Data collected from random subsamples}: HRS measured biomarkers for a randomly selected subsample in 2004 \citep*{hrspm}.  The collection was expanded to a randomly selected half of its sample in 2006 and 2010.  Specimens were collected from the other half of the sample in 2008 and 2012 \citep*{hrsbio08,hrsbio12}.  Incorporating biomarkers as an outcome into the FEM transition model means data for at least half of the sample are missing in any given survey wave.  The four-year spacing between biomarker measurements also introduces the problem of missing data due coarse spacing. \\

\noindent\textbf{Structural changes to survey data over time}:  HRS has been collecting data since the 1991-1992 wave and the FEM typically uses data starting in the 2000 wave for transition model estimation.  Including biomarkers in the FEM transition model means some estimation data are missing from 2000 until the introduction of biomarkers.\\

\noindent\textbf{Item reference time shorter than survey wave spacing}: The PSID interviews collect information about income amounts and sources for the previous calendar year.  Since the interviews are conducted in odd years, the income data are observed only for even years.\\

When data are complete, it is relatively easy to compute the likelihood of the observed outcome sequence and find the maximizing $\hat{\beta}$ value. The likelihood of the sequence is simply the product of the transition probabilities at each time step.
\begin{equation}
\label{eqn:compdatalik}
L(\beta; X_1 \dots X_T) = \prod\limits_{t=2}^T p(X_t \mid X_{t-1}; \beta).
\end{equation}
The marginal probability of $X_1$ is ignored here to keep the focus on transition model estimation. 
The log-likelihood is 
\[
l(\beta; X_1 \dots X_T) = \sum\limits_{t=2}^T \log p(X_t \mid X_{t-1}; \beta).
\]
and the difficulty of finding the maximizer $\hat{\beta}$ depends mostly on the specification of $p(X_t \mid X_{t-1}; \beta).$

When an outcome is missing, the likelihood becomes more difficult to compute because outcomes in the same time step and the next time step depend on the missing outcome. Define $X^\mathrm{(ob)}_t$ as the set of outcomes observed in the survey at time $t$ and $X^\mathrm{(mi)}_t$ as the missing outcomes not observed in the survey at time $t.$
Assume, for simplicity, that $X_1$ is fully observed so that $X^\mathrm{(ob)}_1=X_1$ and $X^\mathrm{(mi)}_1=\emptyset.$
Then, the likelihood of the observed outcome sequence is:
\begin{equation}
\label{eqn:obsdatalik}
L(\beta; X_1 \dots X_T) = \int \dots \int \prod\limits_{t=2}^T  p(X^\mathrm{(ob)}_t\,  X^\mathrm{(mi)}_t \mid X^\mathrm{(ob)}_{t-1}\,  X^\mathrm{(mi)}_{t-1}; \beta) \mathrm{d}X^\mathrm{(mi)}_2 \ldots \mathrm{d}X^\mathrm{(mi)}_T.
\end{equation} 
In this case, the integration makes it more difficult to find the maximizer.  Nevertheless, (\ref{eqn:obsdatalik}) \emph{is} the likelihood of the observed data.  
If the goal is to use maximum likelihood estimates in the simulation phase, then the maximizing $\beta$ value of this likelihood is required.

\subsection{Unified Solution}
This article presents a method to obtain the maximum likelihood estimates from (\ref{eqn:obsdatalik}) that is based on the Expectation-Maximization (EM) algorithm \citep*{dlrem}. Section \ref{sec:theory} shows how the EM algorithm can be applied to estimate Markov Chains using incomplete data and shows how machinery from the simulation phase could be reused to assist in the estimation phase.  The result is an iterative process:
\begin{enumerate}
\item Simulate several copies of the missing survey data using the estimated transition model.
\item Compute imputation weights for each simulated data set.
\item Re-estimate the transition model using the weighted simulated data sets.
\end{enumerate}
This is repeated until the transition model parameter estimates meet some convergence criterion between iterations.
Section \ref{sec:application} demonstrates how to apply this to the FEM and related simulations mentioned in Section \ref{microsims}.
Section \ref{sec:implementation} provides implementation and performance details.  
Although the focus here is on 
solving some common problems in the estimation phase, it turns out that these same tools can be used to solve a related problem in the simulation phase by adjusting the analysis.  This is discussed in Section \ref{sec:discussion} along with suggestions for performance improvements and a note of caution about the types of missing survey data.

\section{EXPECTATION-MAXIMIZATION FOR MARKOV CHAINS}
\label{sec:theory}

\subsection{Expectation-Maximization Algorithm}
The EM algorithm is a general method for maximum likelihood estimation in the presence of missing data.  A brief sketch of EM for a general model is offered to introduce notation before bringing in the additional details that come with Markov chain models.  Suppose a parametric probability model, $p(X; \beta),$ where $X$ is a multivariate random variable.  Suppose further that only some components of $X$ are observed.  Let $X^\mathrm{(ob)}$ and $X^\mathrm{(mi)}$ respectively represent the observed and missing components of $X$.
%The likelihood of $\beta$ given the observed data is $L(\beta; x^\mathrm{(ob)}) = p(x^\mathrm{(ob)}; \beta)$ and the maximum likelihood estimate is $\hat{\beta} = {\arg\max}\, L(\beta; x^{(ob)})$.
%\[
%\hat{\beta} = \underset{\beta}{\arg\max}\, L(\beta; x^{(ob)}) = \underset{\beta}{\arg\max}\, L(\beta; y^{(ob)})
%\]
If $X^\mathrm{(mi)}$ could be observed, then the likelihood of the complete data would be $L(\beta; X^\mathrm{(ob)}\,  X^\mathrm{(mi)}) = p(X^\mathrm{(ob)}\,  X^\mathrm{(mi)}; \beta).$
%\footnote{For simplicity of notation, we assume the arguments to the probability function are ordered with all of the observed outcomes first followed by all of the missing outcomes.}.
  The likelihood for the observed data is an integration over the complete data likelihood with respect to the unobserved data:
\begin{equation}
L(\beta; X^\mathrm{(ob)}) = \int L(\beta; X^\mathrm{(ob)}\,  X^\mathrm{(mi)}) \mathrm{d}X^\mathrm{(mi)}
\label{eqn:gen_complete_lik}
\end{equation}
%where $L\left(\beta; y^{(ob)}, y^{(mi)}\right)$ is the likelihood for both observed and missing data (as if it were observed).  Call $L\left(\beta; y^{(ob)}, y^{(mi)}\right)$ the complete data likelihood.

The EM algorithm provides an iterative approach to finding a maximizer of (\ref{eqn:gen_complete_lik}) instead of trying to find the maximizer directly, which is often difficult because of the integration.  
Start with an initial estimate of the parameters, $\beta^{(0)}$.  Then, repeat the following iteration starting at $k=0:$
\begin{enumerate}
\item \textbf{E-step}: Compute the expected value of the complete data log-likelihood conditional on the observed data and using parameter values $\beta^{(k)}:$
\begin{align}
Q(\beta; \beta^{(k)}) & = E \left[ \log L(\beta; X^\mathrm{(ob)}\,  X^\mathrm{(mi)}) \mid X^\mathrm{(ob)}; \beta^{(k)} \right] \nonumber \\ 
& = \frac{\int \log L(\beta; X^\mathrm{(ob)}\,  X^\mathrm{(mi)}) \,p(X^\mathrm{(ob)}\,  X^\mathrm{(mi)}; \beta^{(k)}) \mathrm{d}X^\mathrm{(mi)}}{\int p(X^\mathrm{(ob)}\,  X^\mathrm{(mi)}; \beta^{(k)}) \mathrm{d}X^\mathrm{(mi)}} \label{eqn:gen_e-step}
\end{align}
%Note a key difference between (\ref{eqn:complete_lik}) and (\ref{eqn:e-step}): In (\ref{eqn:complete_lik}), the same parameter value appears in both the likelihood term, $L\left(\beta; y^{(ob)}, y^{(mi)}\right)$, and the conditional probability for the missing data, $\mathrm{Pr}\left(y^{(mi)} \mid y^{(ob)};  \beta\right)$.  In (\ref{eqn:e-step}), the log-likelihood term depends on $\beta$ while the conditional probability depends on $\beta^{(t)}$.
The second line in (\ref{eqn:gen_e-step}) comes from the fact that the probability of $X^\mathrm{(mi)}$ conditional on $X^\mathrm{(ob)}$ is $p(X^\mathrm{(ob)}\,  X^\mathrm{(mi)}; \beta^{(k)}) / \int p(X^\mathrm{(ob)}\,  X^\mathrm{(mi)}; \beta^{(k)}) \mathrm{d}X^\mathrm{(mi)}$.
Note that $Q(\beta; \beta^{(k)})$ will be used as function of $\beta$, so the E-step is not computing a numerical result, but is instead computing pieces of a function that will be used in the next step.

\item \textbf{M-step}: Keeping $\beta^{(k)}$ constant, find the value of $\beta$ that maximizes $Q(\beta; \beta^{(k)})$ :
\begin{equation}
\beta^{(k+1)} = \underset{\beta}{\arg\max}\, Q(\beta; \beta^{(k)})
\label{eqn:gen_m-step}
\end{equation}

%\item Stop if $\beta^{(k)}$ and $\beta^{(k+1)}$ are similar, e.g., $||\beta^{(t)} - \beta^{(t+1)}|| < 0.00001$ or $$ Otherwise, repeat the iteration using $\beta^{(k+1)}$.
\end{enumerate}
\citet{dlrem} show how the sequence $\beta^{(0)},\beta^{(1)},\beta^{(2)},\ldots$ converges to maximize (\ref{eqn:gen_complete_lik}). In practice, the iteration is typically stopped when some convergence criterion is met, e.g., when $||\beta^{(k+1)} - \beta^{(k)}||$ or $Q(\beta^{(k+1)}; \beta^{(k)}) - Q(\beta^{(k)}; \beta^{(k-1)})$ become very small.
%\subsection{Statistical Inference for Model Parameters}
There are several methods for computing confidence intervals or performing hypothesis tests about the $\beta$ parameters based on the EM estimates. Section 4.2.3 of \citet*{givhot} has an overview of these methods with references.

\subsection{Expectation-Maximization for Estimating Markov Chain Transition Models}
Now, returning to Markov chain models with incomplete estimation data, the complete data likelihood (\ref{eqn:compdatalik}) can be restated in terms of the observed and missing components:
\[
L(\beta; X_1, X^\mathrm{(ob)}_2\,  X^\mathrm{(mi)}_2  \ldots  X^\mathrm{(ob)}_T\,  X^\mathrm{(mi)}_T) = \prod\limits_{t=2}^T  p(X^\mathrm{(ob)}_t\,  X^\mathrm{(mi)}_t \mid X^\mathrm{(ob)}_{t-1}\,  X^\mathrm{(mi)}_{t-1}; \beta).
\]
Taking the log and putting this into the E-step gives:
\begin{align*}
Q(&\beta; \beta^{(k)})  = \\
& \int\dots\int \sum\limits_{t=2}^T \log  p(X^\mathrm{(ob)}_t\,  X^\mathrm{(mi)}_t \mid X^\mathrm{(ob)}_{t-1}\,  X^\mathrm{(mi)}_{t-1}; \beta)
 \,p(X^\mathrm{(mi)}_2 \ldots X^\mathrm{(mi)}_T \mid X_1\,  X^\mathrm{(ob)}_2 \ldots X^\mathrm{(ob)}_T; \beta^{(k)}) \mathrm{d}X^\mathrm{(mi)}_2 \ldots \mathrm{d}X^\mathrm{(mi)}_T
\end{align*}
Applying Bayes' theorem to the term involving $\beta^{(k)}$ yields:
\[
p(X^\mathrm{(mi)}_2 \ldots X^\mathrm{(mi)}_T \mid X_1\,  X^\mathrm{(ob)}_2 \ldots X^\mathrm{(ob)}_T; \beta^{(k)}) = 
\frac{\prod\limits_{t=2}^T  p(X^\mathrm{(ob)}_t\,  X^\mathrm{(mi)}_t \mid X^\mathrm{(ob)}_{t-1}\,  X^\mathrm{(mi)}_{t-1}; \beta^{(k)})}
{\int\dots\int \prod\limits_{t=2}^T  p(X^\mathrm{(ob)}_t\,  X^\mathrm{(mi)}_t \mid X^\mathrm{(ob)}_{t-1}\,  X^\mathrm{(mi)}_{t-1}; \beta^{(k)}) \mathrm{d}X^\mathrm{(mi)}_2 \ldots \mathrm{d}X^\mathrm{(mi)}_T}
\]
The denominator is a function of only $\beta^{(k)}$ and observed data.  It can be ignored when searching for the maximizing $\beta$ value. Putting the numerator back into the E-step gives:
\begin{align}
 Q(&\beta; \beta^{(k)}) = \nonumber \\
& \int\dots\int \sum\limits_{t=2}^T \log  p(X^\mathrm{(ob)}_t\,  X^\mathrm{(mi)}_t \mid X^\mathrm{(ob)}_{t-1}\,  X^\mathrm{(mi)}_{t-1}; \beta)
 \,\prod\limits_{t=2}^T  p(X^\mathrm{(ob)}_t\,  X^\mathrm{(mi)}_t \mid X^\mathrm{(ob)}_{t-1}\,  X^\mathrm{(mi)}_{t-1}; \beta^{(k)}) \mathrm{d}X^\mathrm{(mi)}_2 \ldots \mathrm{d}X^\mathrm{(mi)}_T.
 \label{eqn:mkv_e-step}
\end{align}

\subsection{E-step Computation Using Importance Sampling}
Although the terms inside the integrals in (\ref{eqn:mkv_e-step}) are computationally simple to evaluate, the integrals themselves can be challenging because of the Markov chain dependency structure.  Instead of trying to evaluate these integrals analytically, a Monte Carlo approximation can be used. The ease of this approach depends on the ability to factorize $p(X^\mathrm{(ob)}_t\,  X^\mathrm{(mi)}_t \mid X^\mathrm{(ob)}_{t-1}\,  X^\mathrm{(mi)}_{t-1}; \beta^{(k)})$ into (a) terms that can be easily simulated and (b) terms that can be easily computed.  

One approach is to use the factorization:
\begin{equation}
p(X^\mathrm{(ob)}_t\,  X^\mathrm{(mi)}_t \mid X^\mathrm{(ob)}_{t-1}\,  X^\mathrm{(mi)}_{t-1}; \beta^{(k)}) = p(X^\mathrm{(mi)}_t \mid X^\mathrm{(ob)}_{t-1}\,  X^\mathrm{(mi)}_{t-1}; \beta^{(k)})\,p(X^\mathrm{(ob)}_t \mid X^\mathrm{(mi)}_t\,  X^\mathrm{(ob)}_{t-1}\,  X^\mathrm{(mi)}_{t-1}; \beta^{(k)})
\label{eqn:fac1}
\end{equation}
Starting at $t=2$, where $X_1$ is fully observed, Monte Carlo methods can be used to simulate $R$ draws, $X^\mathrm{(mi)}_{2r}$ ($r=1 \ldots R$), 
from $p(X^\mathrm{(mi)}_2 \mid X_1; \beta^{(k)})$.  Computing $\log p(X^\mathrm{(ob)}_2\,  X^\mathrm{(mi)}_{2r} \mid X_1; \beta)$ and $p(X^\mathrm{(ob)}_2 \mid X^\mathrm{(mi)}_{2r}, X_1; \beta^{(k)})$ for each of these draws gives
\[
\sum \log p(X^\mathrm{(ob)}_2\,  X^\mathrm{(mi)}_{2r} \mid X_1; \beta) p(X^\mathrm{(ob)}_2 \mid X^\mathrm{(mi)}_{2r}\,  X_1; \beta^{(k)}) / R
\] as a Monte Carlo approximation of
\[
\int \log p(X^\mathrm{(ob)}_2\,  X^\mathrm{(mi)}_{2r} \mid X_1; \beta) p(X^\mathrm{(ob)}_2\,  X^\mathrm{(mi)}_2 \mid X_1; \beta^{(k)}) \mathrm{d}X^\mathrm{(mi)}_2.
\]
Moving to $t=3$, Monte Carlo draws of $X^\mathrm{(mi)}_{3r}$ can be simulated from 
$p(X^\mathrm{(mi)}_3 \mid X^\mathrm{(ob)}_2\,  X^\mathrm{(mi)}_{2r}; \beta^{(k)})$ for $i=1 \ldots R$ and then compute $p(X^\mathrm{(ob)}_3 \mid X^\mathrm{(mi)}_{3r}\,  X^\mathrm{(ob)}_2\,  X^\mathrm{(mi)}_{2r}; \beta^{(k)})$.  Combining the time steps gives
\[
\sum\limits_{r=1}^R \sum\limits_{t=2}^3 \log  p(X^\mathrm{(ob)}_t\,  X^\mathrm{(mi)}_{tr} \mid X^\mathrm{(ob)}_{t-1}\,  X^\mathrm{(mi)}_{(t-1)r}; \beta)\prod\limits_{t=2}^3  p(X^\mathrm{(ob)}_t \mid X^\mathrm{(mi)}_{tr} , X^\mathrm{(ob)}_{t-1}, X^\mathrm{(mi)}_{(t-1)r}; \beta^{(k)}) / R
\]
 as a Monte Carlo approximation of
\[
\int\int \sum\limits_{t=2}^3 \log  p(X^\mathrm{(ob)}_t\,  X^\mathrm{(mi)}_{tr} \mid X^\mathrm{(ob)}_{t-1}\,  X^\mathrm{(mi)}_{(t-1)r}; \beta)\prod\limits_{t=2}^3  p(X^\mathrm{(ob)}_t \mid X^\mathrm{(mi)}_{tr} \,  X^\mathrm{(ob)}_{t-1}\,  X^\mathrm{(mi)}_{(t-1)r}; \beta^{(k)}) \mathrm{d}X^\mathrm{(mi)}_2 \mathrm{d}X^\mathrm{(mi)}_3.
\]
Continuing like this through time $T$, let
\[
w_r(\beta^{(k)}) = \prod\limits_{t=2}^T  p(X^\mathrm{(ob)}_t \mid X^\mathrm{(mi)}_{tr} \,  X^\mathrm{(ob)}_{t-1}\,  X^\mathrm{(mi)}_{(t-1)r}; \beta^{(k)}),
\]
then $\sum_{r=1}^R \sum_{t=2}^T \log  p(X^\mathrm{(ob)}_t\,  X^\mathrm{(mi)}_{tr} \mid X^\mathrm{(ob)}_{t-1}\,  X^\mathrm{(mi)}_{(t-1)r}; \beta) w_r(\beta^{(k)})$ 
is a Monte Carlo estimate of  (\ref{eqn:mkv_e-step}).  
The goal here is to simulate from the distribution of missing outcomes conditional on all observed outcomes in past, present, and future time steps.  This is accomplished by simulating the missing outcomes from a different distribution (dependent only on the previous time step), then weighting the simulated outcomes to match the target distribution (dependence on current and future time steps).

If (\ref{eqn:fac1}) does not yield convenient simulation and computation, other factorizations can be considered, such as,
\begin{equation}
p(X^\mathrm{(ob)}_t\,  X^\mathrm{(mi)}_t \mid X^\mathrm{(ob)}_{t-1}\,  X^\mathrm{(mi)}_{t-1}; \beta^{(k)}) = p(X^\mathrm{(mi)}_t \mid X^\mathrm{(ob)}_t\,  X^\mathrm{(ob)}_{t-1}\,  X^\mathrm{(mi)}_{t-1}; \beta^{(k)})\,p(X^\mathrm{(ob)}_t \mid X^\mathrm{(ob)}_{t-1}\,  X^\mathrm{(mi)}_{t-1}; \beta^{(k)})
\label{eqn:fac2}
\end{equation}
Here, $X^\mathrm{(mi)}_t$ is simulated conditional on the the previous time step \emph{and} observed data in the same time step, $X^\mathrm{(ob)}_t$.  Repeating $R$ times and letting
\[
w_r(\beta^{(k)}) = \prod\limits_{t=2}^T  p(X^\mathrm{(ob)}_t \mid X^\mathrm{(ob)}_{t-1}\,  X^\mathrm{(mi)}_{(t-1)r}; \beta^{(k)}),
\]
gives $\sum_{r=1}^R \sum_{t=2}^T \log  p(X^\mathrm{(ob)}_t\,  X^\mathrm{(mi)}_{tr} \mid X^\mathrm{(ob)}_{t-1}\,  X^\mathrm{(mi)}_{(t-1)r}; \beta) w_r(\beta^{(k)})$
as the Monte Carlo approximation of  (\ref{eqn:mkv_e-step}).  

If outcomes in the same time step are conditionally independent given outcomes at the previous time step, then the factorization is simply:
\begin{equation}
p(X^\mathrm{(ob)}_t\,  X^\mathrm{(mi)}_t \mid X^\mathrm{(ob)}_{t-1}\,  X^\mathrm{(mi)}_{t-1}; \beta^{(k)}) = p(X^\mathrm{(mi)}_t \mid X^\mathrm{(ob)}_{t-1}\,  X^\mathrm{(mi)}_{t-1}; \beta^{(k)})\,p(X^\mathrm{(ob)}_t \mid X^\mathrm{(ob)}_{t-1}\,  X^\mathrm{(mi)}_{t-1}; \beta^{(k)})
\label{eqn:facci}
\end{equation}
Simulate $X^\mathrm{(mi)}_{tr}$ from $p(X^\mathrm{(mi)}_t \mid X^\mathrm{(ob)}_{t-1}\,  X^\mathrm{(mi)}_{t-1}; \beta^{(k)})$ and compute weights, 
\[w_r(\beta^{(k)}) = \prod\limits_{t=2}^T p(X^\mathrm{(ob)}_t \mid X^\mathrm{\tiny{(ob)}}_{t-1}\,  X^\mathrm{(mi)}_{(t-1)r}; \beta^{(k)}).
\]  Due to the conditional independence assumption in the transition model, the ability to simulate from $p(X^\mathrm{(mi)}_t \mid X^\mathrm{(ob)}_{t-1}\,  X^\mathrm{(mi)}_{t-1}; \beta^{(k)})$ will also be needed for the simulation phase.
As a result, conditional independence allows the simulation developer to solve two problems with one implementation.  It is not necessary to use the same factorization in every time step.  As shown in Section \ref{sec:application}, the FEM uses conditional independence (\ref{eqn:facci}) for all outcomes when an individual is alive at time $t,$ factorization (\ref{eqn:fac1}) when the mortality status is missing, and (\ref{eqn:fac2}) if the individual was alive at time $t-1$ and died by time $t$.

In practice, without conditional independence, the factoring of outcomes into a missing block and an observed block in (\ref{eqn:fac1}) and (\ref{eqn:fac2}) might not lead immediately to a structure that is easily simulated or weights that are easily computed.  In that case, it is necessary to look at the dependencies between individual outcomes in $X^\mathrm{(ob)}_t$ and $X^\mathrm{(mi)}_t$ and to factorize into smaller blocks of outcomes.  This is completely dependent on the transition model structure in a particular simulation project, but the guiding principle is to look for blocks of missing outcomes that can be easily simulated by using the same functions needed for the simulation phase.

To summarize, the EM algorithm is modified by inserting imputation (I) and weighting (W) steps:
\begin{enumerate}
\item \textbf{E-step}: For $r=1 \ldots R$:
\begin{enumerate}
\item[1.a.] For $t=2 \ldots T$: 
\begin{enumerate}
\item[1.a.1.] \textbf{I-step}: Simulate the missing outcomes at time $t$ to create a complete data set: $X_{tr}$
\item[1.a.2.] \textbf{W-step}: Compute the probability of the observed outcomes at time $t$: $w_{rt}(\beta^{(k)})$ 
\end{enumerate}
\item[1.b.] Compute $w_{r}(\theta^{(k)}) = \prod\limits_{t=2}^T w_{rt}(\beta^{(k)})$ 
%\footnote{In fact, one only needs to compute the probabilities for the observed outcomes at time t which depend on imputed outcomes at time (t-1). Probabilities for observed outcomes at time t which depend only on observed outcomes at time (t-1) are constant across multiple imputations. They can be factored out of the numerator and denominator(?) integrals in (EQN) to cancel each other out.}
\end{enumerate}
\item \textbf{M-step}: Keeping $\beta^{(k)}$ constant, find 
\[
\beta^{(k+1)} = \underset{\beta}{\arg\max}\, \sum_{r=1}^R \sum_{t=2}^T \log  p(X_{tr} \mid X_{(t-1)r}; \beta) w_r(\beta^{(k)})
\]
\item Stop when convergence criterion is satisfied.
\end{enumerate}
The implementation details of the I-step and W-step depend on how $p(X^\mathrm{(ob)}_t\, X^\mathrm{(mi)}_t \mid X^\mathrm{(ob)}_{t-1}\,  X^\mathrm{(mi)}_{t-1}; \beta^{(k)})$ is factorized at each time $t.$  The convergence check is required as the final step because using finite $R$ will leave some Monte Carlo error between $\beta^{(k)}$ and $\beta^{(k+1)}$. This approach can be seen as a hybrid between the analytical E-step of \citet{dlrem} and the fully Monte Carlo E-step of \citet*{mcem}. This method is called EMMCIS (Expectation-Maximization for Markov Chains using Importance Sampling) in what follows.

\section{APPLICATION TO FUTURE ELDERLY MODEL}
% FAMILY OF MICROSIMULATIONS
\label{sec:application}

The FEM simulates a rich set of outcomes from HRS, including physical and mental health status, disability, health risk factors, employment status, earnings, retirement income, and Medicare and Medicaid participation.  
A simplified version of the model is used here to demonstrate the new method.  
It captures important features of the FEM, but sacrifices richness for ease of demonstration.  
Specifically, the simplified model includes current smoking status, diagnosis of diseases, and mortality as binary yes/no transitioned outcomes. 
Age is computed from birth year. Gender and race/ethnicity are fixed characteristics. 
The possible disease diagnoses are cancer, diabetes, heart disease, hypertension, lung disease, and stroke.  
Each disease diagnosis is an absorbing state -- if an individual reports a diagnosis at time $t$ they will remain in that state for every subsequent time step. 
Smoking status is transient.  
The demographic variable states are male or female for gender and white, Hispanic, or non-Hispanic black for race/ethnicity.  
Following the FEM, the simplified model is a first-order Markov chain and outcomes are independent conditional upon the state in the previous time step and mortality in the current time step.  
If an individual dies in the current time step, then the model does not simulate their transition in any of the other outcomes. 
This matches the structure of HRS core interview data used for estimation \citep*{rndhrsp}.

\begin{table}[htb]
\centering
\caption{Dependencies between transitioned outcomes in the simplified FEM model. All outcomes also depend on age and demographics.}
\begin{tabular}{lcl}
\hline
Outcome && Dependencies in previous time step \\\hline
Smoking status && cancer, diabetes, heart disease, hypertension, lung disease, stroke, smoking status\\
Cancer && smoking status\\
Diabetes && smoking status\\
Heart disease && diabetes, hypertension, smoking status\\
Hypertension && diabetes, smoking status\\
Lung disease && smoking status\\
Stroke && cancer, diabetes, heart disease, hypertension, smoking status \\
Mortality && cancer, diabetes, heart disease, hypertension, lung disease, stroke, smoking status\\
\hline
\end{tabular}
\label{tab:depstruct}
\end{table}

Each of the FEM outcomes considered here is modeled as a Probit regression.  Let the outcomes for individual $i$ at time $t$ be denoted as $x_{ijt}$  with $x_{i1t}$ as smoking status and $x_{i8t}$ as mortality status.  The outcomes $x_{ijt}$ for $j=2 \ldots 7$ are the six disease conditions, but the exact assignment of disease condition to $j$ index is not important.  Let $D_{ijt}$ be the vector of outcomes at time $t-1$ on which $x_{ijt}$ depends (see Table \ref{tab:depstruct}).  For example, $D_{i1t} = [x_{i1(t-1)}\, x_{i2(t-1)}\, \dots \,x_{i8(t-1)}]$ because smoking status at time $t$ depends on all outcomes in the previous time step. Let $z_{ijt}$ be the vector of individual covariates for outcome $j$ at time $t,$ containing a unit constant, age, demographic indicators, and the elements of $D_{ijt}.$  The Probit model for each outcome has its own parameter vector $\beta_j$ of coefficients corresponding to the elements of $z_{ijt}$: an intercept term, an age coefficient, coefficients for demographics, and coefficients for outcomes in the previous time step.
Each transition probability is then computed as $\Phi(z_{ijt}^T \beta_{j})$ with adjustments for absorption and mortality, where appropriate.  
Let $X_{it}$ be the vector of outcomes, including age and demographics, for individual $i$ at time $t.$  The log-likelihood has the following components:
\[
l_{i1t}(\beta_1; X_{i(t-1)}\, X_{it}) = (1-x_{i8t})\left( x_{i1t} \log \Phi(z_{i1t}^T \beta_1) + (1-x_{i1t}) \log (1-\Phi(z_{i1t}^T \beta_1)) \right)
\]
for smoking status,
\[
l_{ijt}(\beta_j; X_{i(t-1)}\, X_{it}) = (1-x_{i8t})(1-x_{ij(t-1)})\left(x_{ijt} \log \Phi(z_{ijt}^T \beta_j) + (1-x_{ijt}) \log (1-\Phi(z_{ijt}^T \beta_j))\right)
\]
for disease conditions ($j=2 \ldots 7$), and
\[
l_{i8t}(\beta_8; X_{i(t-1)}\, X_{it}) = (1-x_{i8(t-1)})\left(x_{i8t} \log \Phi(z_{i8t}^T \beta_8) + (1-x_{i8t}) \log (1-\Phi(z_{i8t}^T \beta_8))\right)
\]
for mortality.
Putting everything together, the log-likelihood for the simplified FEM model is:
\[
l(\beta; X_1 \ldots X_T) = \sum_{i=1}^n \sum_{t=2}^{T} \sum_{j=1}^8 l_{ijt}(\beta_j; X_{i(t-1)}\,X_{it})
\]
This is the likelihood of the model to be \emph{simulated}.  However, the HRS data for estimating it are incomplete.

There are missing responses in HRS data because some respondents could not be contacted at all in a particular survey wave or because some respondents did not answer specific questions.  For the sake of demonstration, assume all of these data are missing at random conditional on the observed data.  See Section \ref{sec:mar} for discussion.  Additionally, the model uses a one-year time step.  This means HRS data are only available for every other time step.  However, the missing time steps can be filled in with information about absorbing outcomes because $x_{ijt} = 1$ implies $x_{ijt'} = 1$ for all $t' > t$ and $x_{ijt} = 0$ implies $x_{ijt'} = 0$ for all $t' < t.$  The only times when $x_{ijt}$ will be missing for absorbing outcomes are when the most recent observation is 0 and the subsequent observation is 1. Individuals' ages and demographics are always observed.  The log-likelihood for the observed data is:
\[
l(\beta; X^{(ob)}_1 \ldots X^{(ob)}_T) = \sum_{i=1}^n \log \int \dots \int \prod\limits_{t=2}^T \prod\limits_{j=1}^8 \exp\left\{l_{ijt}(\beta_j; X^{(ob)}_{t-1}\, X^{(mi)}_{t-1}\, X^{(ob)}_t\, X^{(mi)}_{t})\right\}  \mathrm{d}X^{(mi)}_{i2} \ldots \mathrm{d}X^{(mi)}_{iT},
\]
where $X^{(ob)}_{it}$ and $X^{(mi)}_{it}$ are the observed and missing outcomes for individual $i$ at time $t$, respectively. 
%For even $t$ we have $X^{(ob)}_{it} = \emptyset.$

%\subsection{Motivations for Shorter Time steps} [DOES THIS GO IN INTRO?]
%- lifetime cost calculation with 1-year cost models
%- implementing interventions with 1-year follow-up

The EMMCIS method was implemented for this demonstration by iterating over individuals at the highest level and iterating sequentially over time steps within each individual.  For each time step, the first task is to check mortality status and impute it if it is unknown.  Then, if the individual is alive at time $t$ (either observed or imputed),  the conditional independence of outcomes allows processing the remaining outcomes in any order or in parallel.
The E-step is applied to each individual $i$ is as follows:
\begin{enumerate}
\item For $r=1 \ldots R$:
\begin{enumerate}
\item[1.a.] Initialize $w'_{ir} = 0$
\item[1.b.] For $t=2 \ldots T$: 
\begin{enumerate}
\item[1.b.1.] If $x_{i8t}$ is missing, simulate $x_{i8tr}=1$ with probability $\Phi(z_{i8tr}^T \beta^{(k)}_8)$, otherwise set $x_{i8tr} = x_{i8t}$ and update $w'_{ir} = w'_{ir} + l_{i8t}(\beta_j^{(k)}; X_{i(t-1)r}\, X_{it})$
\item[1.b.2.] If $x_{i8tr} = 1$, stop iteration over $t$ and move to next $r$ in step 1.
\item[1.b.3.] For $j=1 \ldots 7$:
\begin{enumerate}  
\item[1.b.3.a.]  If $x_{ijt}$ is missing, simulate $x_{ijtr}=1$ with probability $\Phi(z_{ijtr}^T \beta^{(k)}_j)$, otherwise set $x_{ijtr} = x_{ijt}$ and update $w'_{ir} = w'_{ir} + l_{ijt}(\beta_j^{(k)}; X_{i(t-1)r}\, X_{it})$
\end{enumerate}
\end{enumerate}
\item[1.c.] Compute $w_{ir} = \exp(w'_{ir})$ 
\end{enumerate}
\end{enumerate}
Note that the I-step and W-step are applied to each outcome in steps 1.b.1 and 1.b.3.a.  The M-step is computed by combining imputations across individuals and the weights:
\begin{equation}
\beta^{(k+1)} = \underset{\beta}{\arg\max}\, \sum_{i=1}^n \sum_{r=1}^R \sum_{t=2}^T \sum_{j=1}^8 l_{ijt}(\beta_j; X_{i(t-1)r}\, X_{itr}) w_{ir}.
\label{eqn:app_m-step}
\end{equation}

\section{IMPLEMENTATION}
\label{sec:implementation}

The performance of EMMCIS in estimating the simplified FEM model was evaluated using $n=1010$ individuals surveyed by HRS in 2000 and followed until death or 2014.  
%Only individuals who were fully observed in the 2000 survey wave were included to avoid the problem of imputing missing data at time $t=1.$  
The estimation was implemented in R \citep*{rcore} and run on a server using 20 CPUs for parallelization over individuals.  The server had 196 GB of memory and 2.4 GHz clock speed for each CPU.  There was some competition for CPUs from other users while the estimation was running, but the estimation had at least 85\% of the CPU time.

The R implementation used the \emph{plyr} package \citep*{plyr} with the \emph{doParallel} backend \citep*{dopar} to divide work between nodes.  
To generate an initial estimate of each $\beta_j$, a Probit regression was run on the observed data, i.e., treating two-year time steps as a one-year time step and throwing away any transitions with a missing outcome value ($x_{ijt}$) or missing values for dependencies in the previous time step ($D_{ijt}$).  The M-step used the BFGS method for optimization via the \emph{optim} function in R.  The BFGS iteration was run for a pre-specified number of iterations or until its default convergence criterion (a relative likelihood change of less than roughly $10^{-8}$) was met.  The EM iteration was stopped when the absolute change in likelihood between subsequent iterations was less than $10^{-4}.$  

There are two conceptual gains in each EM iteration. First, there is a gain from computing the E-step with a $\beta^{(k)}$ value that is closer to the maximizer than $\beta^{(k-1)}$.  Second, there is a gain from running the M-step to find a more optimal $\beta$ value.  Each of these gains has a cost: the E-step requires simulating $R$ copies of the missing values and the M-step involves repeatedly evaluation the $Q(\beta; \beta^{(k)})$ function (and its gradient).  During implementation tests, it was observed that initial EM iterations made large gains of the second kind from only one BFGS iteration while the first kind of gain had smaller magnitudes.  This was true even when $R$ was small.  This led to an adaptive iteration scheme where the number of iterations in the E-step and M-step were increased based on changes in the $Q(\beta; \beta^{(k)})$ function.  At the beginning of the estimation, $R$ was set to 10 and the maximum number of BFGS iterations was set to 3.  If any iteration ended with 
$Q(\beta^{(k+1)}; \beta^{(k)}) - Q(\beta^{(k)}; \beta^{(k-1)}) \leq 1.97\times10^{-4} \times \sqrt R$ then $R$ was increased by a factor of ten.  The threshold for increasing the number of iterations was based on a roughly estimated upper bound for the Monte Carlo variance of $Q(\beta; \beta^{(k)})$ using a small number of Monte Carlo replicates. Further work is needed to determine an appropriate threshold.  After $R$ reached 1,000, then, in any iteration that failed to increase $Q(\beta^{(k+1)}; \beta^{(k)})  - Q(\beta^{(k)}; \beta^{(k-1)})$ beyond the threshold, the maximum number of BFGS iterations was increased by a factor of 10 until reaching 300. With this adaptive strategy, the estimation converged on the 13th EM iteration after running for 126 hours. Across all EM iterations,  $Q(\beta; \beta^{(k)})$ was evaluated 998 times and its gradient 112 times. 

\section{DISCUSSION}
\label{sec:discussion}

\subsection{Paths to Improved Performance}
There are some obvious ways to make the estimation in Section \ref{sec:implementation} use computational resources more efficiently.  As a first step, some efficiency will be gained by translating the R code to a compiled language like C$++$.  However, the greatest gains should come from minimizing the amount of data communicated between processing nodes and reducing the number of iterations in the M-step optimization. Implementation of any of these performance improvements is expected to make the estimation to run noticeably faster.

The current implementation starts by loading all of the observed data into the master node's memory.  During the E-step, the \emph{plyr} methods in R communicate pieces of the observed data set to each worker node and the worker nodes return the corresponding complete data imputations.  Pieces of the complete data imputations are then sent back to the worker nodes during each optimization iteration in the M-step.  It is possible to eliminate the communication of individual data points in a distributed memory environment with shared storage by creating worker processes with the following functionality.
\begin{itemize}
\item \textbf{Initialization}: Load a section of the observed data from storage into local memory, e.g., the $p$th node of $P$ total nodes will load data for $n/P$ individuals.
\item \textbf{E-step}: Receive $\beta^{(k)}$ from master node.  For each individual stored locally, create $R$ imputations of missing data and compute $w_{ir}$.
\item \textbf{M-step}: Receive $\beta$ from master node. Evaluate the complete data likelihood component (or its gradient or Hessian) for each individual stored locally. Send the weighted sum over these individuals back to the master node.
\end{itemize}
With this functionality, the largest object communicated between processes is a gradient or Hessian, both of which are much smaller than the observed and imputed data sets.

There are several ways that the M-step implementation can be improved.  The overall goal is to reduce the number of optimization iterations because evaluation of the $Q(\theta; \theta^{(k)})$ function or its derivatives is very expensive.  The \emph{optim} function in R evaluates $Q(\theta; \theta^{(k)})$ and its gradient separately.  However, the same iteration over individuals, imputations, and time steps is required for both evaluations, so it would be more efficient to compute $Q(\theta; \theta^{(k)})$ and its gradient during the same iteration when both evaluations are needed.
Next, efficiency can be improved by exploiting the independence of the model parameters. Because the $\beta_j$ vectors are specific to each outcome, the M-step in (\ref{eqn:app_m-step}) can be separated into eight independent maximization problems:
\[
\beta_j^{(k+1)} = \underset{\beta_j}{\arg\max}\, \sum_{i=1}^n \sum_{r=1}^R \sum_{t=2}^T  l_{ijt}(\beta_j; X_{i(t-1)r}\, X_{itr}) w_{ir}.
\]
Finally, with this independence in mind, Newton's Method might be a faster alternative to the BFGS method.  The FEM models rarely have more than 20 coefficient parameters in each $\beta_j$ and it is relatively easy to compute the Hessian of each likelihood component.  

\subsection{Appropriate Types of Missingness}
\label{sec:mar}

The incomplete data examples discussed in \ref{sec:incdata} are all due to circumstances unrelated to the state of the survey respondent.  In general, the method developed here can be applied when the unobserved state of a missing outcome is independent of the fact that its missing, conditional on all other observed data.  The method cannot be applied to a missing data point for an individual when combining the observed data for that individual with the fact the the data point is missing makes the unobserved state more or less likely.  In these cases, the transition likelihood needs to be adjusted to condition on the missing status of the outcome.
%For example, there is some probability $p$ that a survey respondent will have a disease diagnosis at time $t$ given that they are healthy at time $t-1.$  If the individual does not participate in the survey at time $t,$ then it might be more likely that they have disease diagnosis at time $t$. 
%
%	R	NR
%S
%H
\citet*{little_rubin} provide a detailed discussion of the EM algorithm applied to different missing data mechanisms.

\subsection{Comparison of Other Methods}
A maximum likelihood solution to the coarse data spacing issue discussed in Section \ref{sec:incdata} is given by \citet*{craig} and, more recently, \citet*{factoring_orig}. \citet{craig} also provide an EM solution to the irregular data spacing issue.  The method presented here is more broadly applicable because it does not require discrete states and does not require homogeneity or stationarity assumptions.  The present method can also be applied more generally because it does not depend on any structural pattern in the observed and missing data.

\subsection{Connection to Bridge Simulations}
\label{sec:bridgesim}
The EMMCIS method addresses the problem of incomplete survey data during the estimation phase. There is a related problem in the simulation phase that can be solved using the same tools.  
Suppose that a fully specified transition model is already present (either estimated or obtained from some external source) and the goal is to simulate outcomes starting with observed outcomes $X_1$ as the initial state. This is straightforward: use the transition model to simulate $X_2$ conditional on the observed $X_1$ outcomes, simulate $X_3$ conditional on the simulated $X_2$ outcomes, and so on.  Now suppose that outcomes from a later time are also observed as $X_T$ with $T>2$.  How can the outcomes $X_2 \ldots X_{T-1}$ be simulated from the given transition model, but conditional on the observed starting and ending states, $X_1$ and $X_T$?  Following Section \ref{sec:theory}, the target model for simulation is:
\[
\Pr(X_2 \ldots X_{T-1} \mid X_1\,  X_T; \beta) = \frac{p(X_2 \mid X_1; \beta)\dots p(X_{T} \mid X_{T-1}; \beta)}
{\int\dots\int p(X_2 \mid X_1; \beta) \dots p(X_{T-1} \mid X_{T-2}; \beta) p(X_{T} \mid X_{T-1}; \beta) \mathrm{d}X_2 \ldots \mathrm{d}X_{T-1}}
\]
Once again, the model can be simulated by importance sampling.  After simulating a replicate of the chain, $X_{2m} \ldots X_{(T-1)m},$ using the transition model, compute the replicate weight as $w_m = p(X_{T} \mid X_{(T-1)m}; \beta)$.  Results in the analysis phase are then computed as a weighted average: 
\[
\hat{T} = \sum_{m=1}^M T(X_1\,  X_{2m} \ldots X_{(T-1)m}\,  X_T) w_m / \sum_{r=1}^M w_m.
\]
Section 6.4.1 of \citet{givhot} discusses bias and other important theoretical considerations when using importance weights in the analysis.

\section*{ACKNOWLEDGMENTS}
The HRS (Health and Retirement Study) is sponsored by the National Institute on Aging (grant number NIA U01AG009740) and is conducted by the University of Michigan. This work was supported by the National Institute On Aging of the National Institutes of Health under Award Number P30AG024968. The content is solely the responsibility of the author and does not necessarily represent the official views of the National Institutes of Health.

\bibliographystyle{apalike}
\bibliography{fem_em}

\end{document}